\documentclass[varenna]{cimento}

\usepackage{amsmath,amssymb,float,rotating}
\usepackage{graphicx}  
\usepackage{graphics, setspace}

\title{A primer on statistically validated networks}

\author{Salvatore~Miccich\`e~$^a$
\atque
Rosario~N. Mantegna$^{a,b,c}$}

\institute{(a) Dipartimento di Fisica e Chimica, Universit\`a di Palermo, Viale delle Scienze, Ed. 18, I-90128, Palermo, Italy\\
(b) Complexity Science Hub Vienna, Josefst\"adter Stra\ss e 39, 1080 Vienna, Austria\\
(c) Computer Science Department, University College London, Gower Street, London, WC1E 6BT, UK}

\shortauthor{S. Miccich\`e \atque R.~N. Mantegna}

\begin{document}

\maketitle



\begin{abstract}
In this contribution we discuss some approaches of network analysis providing information about single links or single nodes with respect to a null hypothesis taking into account the heterogeneity of the system empirically observed. With this approach, a selection of nodes and links is feasible when the null hypothesis is statistically rejected. We focus our discussion on approaches using (i) the so-called disparity filter and (ii) statistically validated network in bipartite networks. For both methods we discuss the importance of using multiple hypothesis test correction. Specific applications of statistically validated networks are discussed. We also discuss how statistically validated networks can be used to (i) pre-process large sets of data and (ii) detect cores of communities that are forming  the most close-knit and stable subsets of clusters of nodes present in a complex system.
\end{abstract}

\section{Introduction}
Network science is a multidisciplinary research area investigating complex networks \cite{Newman,Havlin,Barabasi,Latora} in complex systems. Networks are present in many complex systems of different origin and nature. Examples are biological,  physiological, ecological, economic, social and financial networks.  

In network science networks are characterized at different hierarchical levels. In many cases, the most basic characterization concerns the type of degree distribution observed. In fact the type of degree distribution strongly affects the properties of the overall network with respect to key problems such as, for example, the spreading of epidemics or the robustness of the network to failures and attacks.  Another level of analysis concerns the local structures observed in complex networks. In fact a widespread characteristic of complex network is the presence of subnetworks of nodes with internal interconnections more intense than what expected on the basis of the degree heterogeneity observed in the overall network. These subnetworks are called "communities" and their statistical detection is often referred to as "community detection" because the first observation of such subnetworks was in the pioneering studies of social networks performed in the seventies of last century. A classic example of these type of study is the famous antropological study of the Zachary's Karate club \cite{Zachary}.

Essentially most real complex networks present  an organization of network in interconnected sub-units (i.e. the sub-units customary called ``communities" or, more generally clusters of nodes). This organization in sub-units is typically highly informative about the investigated network \cite{Girvan}. This observation has triggered a huge interest towards the development of accurate, performing and informative ``community detection algorithm" and communities detected in real complex networks have been characterized and interpreted in many real systems \cite{Fortunato}.

The level of communities is not the most refined level in a network. In fact, at a more refined level one can consider the statistical properties of so-called $k$-motifs. These structures are small subnetworks each involving $k$ nodes. Social scientists have studied 3-motifs and 4-motifs since the seventies of the last century \cite{Wasserman}. 
More recently many other disciplines including system biology, computer science, statistical physics and computational social science have also investigated the over-representation and under-representation of $k$-motifs with respect to a null hypothesis. One of the first example of these approach is presented in \cite{Alon}

Both community detection and the investigation of $k$-motifs uses a random null hypothesis to detect communities or to evaluate whether the observation of a certain number of a specific $k$-motif is over-expressed or under-expressed (with respect to the null hypothesis.  A null model is constructed by considering the expected distribution of a given quantity.  An empirical quantity can therefore be compared with this distributional expectation and this comparison is providing a tool to perform a statistical test aiming at verifying whether  a given empirical value is statistically compatible with the null hypothesis or rather reject it. When the null hypothesis is rejected empirical data are associated with aspects and properties that are not taken into account in the considered null model. Coming back to the search of $k$-motifs, the observation of over-representation (or under-representation) of some 3-motifs turned out to be highly informative in social networks for the interpretation of social interactions (see, for example, ref.  \cite{Wasserman}).

The most refined level of analysis of a network concerns the statistical validation of of single nodes or single links. Starting from 2009 some studies have considered the over-representation or under-representation of a property of a node and/or of a link.  In other words, with this approach one is able to highlight statistical deviations from a null hypothesis at the level of a single node or at the level of a pair of nodes connected by a link. These approaches are relatively recent. For this reason, there is still no consensus on the general term to be used to address this approach. The names proposed so far are (i) backbone of a network \cite{Serrano}, and (ii) statistically validated network \cite{Tumminello2011}. Selection of nodes or links not consistent with a null hypothesis have been investigated in studies focusing on classic examples of networks \cite{Serrano,Tumminello2011}, trading decisions of investors \cite{Tumminello2012,Musciotto2016,Musciotto2018,Challet}, criminal career of a large set of suspects  \cite{Tumminello2013}, mobile phone calls of large set of users \cite{Li2014a,Li2014b}, financial credit transactions occurring in an Interbank market \cite{Hatzopoulos}, intraday lead-lag relationships of returns of financial assets traded in major financial markets \cite{Curme}, the Japanese credit market \cite{Marotta}, the socio-technical system of air traffic management \cite{BongiornoFork}, the core of communities observed in projected networks originating from bipartite networks \cite{BongiornoCore}, the international trade \cite{Garlaschelli}, and temporal social ties observed in face to face interactions \cite{Barrat}.

In this paper, we present a primer on how to highlight single nodes and/or single links in a complex network by selecting a null hypothesis and by performing a large number of statistical tests on all nodes or links of the network. Nodes and links are highlighted when the null hypothesis is statistically rejected. This approach usually requires a large number of statistical tests. In the presence of a large number of statistical tests a multiple hypothesis test correction is needed to avoid the presence of false positive. In this paper we discuss the importance of multiple hypothesis test correction and the properties and limitations of the most used corrections. The paper is organized as follows. In Section 2 we discuss the so-called disparity filter that was the first procedure introduced to select a subset of links of a complex network based on a statistical test. Section 3 discusses the need of a multiple hypothesis test correction in the process of statistical validation of all links belonging to a complex network.Section 4 describes the method used to obtain statistically validated networks in bipartite systems. Section 5 presents some examples and types of applications of statistically validated networks. Section 6 discusses community detection in statistically validated networks whose projected network originates from a bipartite network. Section 7 provides information about a few software packages detecting statistically validated networksand Section 8 presents some conclusions.

\section{Disparity filter}
To the best of our knowledge, the filtering procedure introduced in \cite{Serrano} for weighted networks was the first method used to highlight nodes and links of a complex network without using a thresholding approach. In fact the method performs statistical tests on all elements of the networks and selects all nodes and links that reject a specific null hypothesis. 
The statistical test is performed on the weight node $i$ has with node $j$. The statistical test performed are therefore tests involving pairs of nodes and their link.

Let us describe briefly the method. The analyzed network is a weighted network. Let $k_i$ and $s_i$ be respectively the degree and the strength of node $i$, and let $w_{ij}$ be the weight of the link observed between node $i$ and $j$. We set $x_{ij}=w_{ij}/s_i$, i.e.  $x_{ij}$ is the normalized weight of the link with respect to node $i$.

The null hypothesis used in \cite{Serrano} assumes that node $i$ distribute its total strength in a uniform random division among all connected nodes. 

To visualize this process for a node of degree $k$, let us consider $k-1$ points distributed with uniform probability in the interval $[0, 1]$ so that it ends up divided into $k$ subintervals. Their lengths would represent the expected values for the $k$ normalized weights $x_{ij}$ according to the null hypothesis. The the probability density function $\rho(x_{ij})$ to observe an interval with length $x_{ij}$ in the $k$ sub-intervals dividing the interval $[0, 1]$. 
\begin{equation}
\rho(x_{ij})=(k_i-1)(1-x_{ij})^{k_i-2}
\end{equation}
By using this probability one can obtain the p-value associated with the observation of a value $x_{ij}$ or smaller is 
\begin{equation}
 \label{weight_prob}
 p(x_{ij})=1-(k_i-1)\int_{0}^{x_{ij}}(1-x_{ij})^{k_i-2}dx.
\end{equation}

If $p(x_{ij})$ in Eq. \ref{weight_prob} is smaller than a given, predetermined, statistical threshold $\theta$ the null hypothesis is rejected and the link between node $i$ and node $j$ is selected to be part of the statistically validated network (called backbone in the original paper). 
It is worth noting that a statistical test is performed on each node $i$ and each link between $i$ and $j$ is tested both from the perspective of node $i$ and from the perspective of node $j$. Therefore, the backbone obtained is intrinsically a directed network.  The disparity filter assumes as a null hypothesis equal distribution of the strength of a node to all connected nodes. This is a quite restrictive assumption because in many real complex systems the strength of a node is not shared uniformly divided with the connected nodes but rather a significant heterogeneity among the different weights is observed. This point is discussed in  \cite{Radicchi2011} where a generalized approach to the case of non uniform distribution of weights is proposed.

Another aspect to be cited is that the number of statistical tests to be performed is usually very large and therefore it is necessary to perform a multiple hypothesis test correction to avoid the presence of a large number of false positive. This type of correction was not present in the original papers \cite{Serrano,Radicchi2011} but it has been used in a subsequent application of the methodology \cite{Marotta}. In the next section we discuss the necessity of the multiple hypothesis test correction, the methods used to perform the correction, their strength and their limitations.  

\section{Multiple hypothesis test correction}
Let us first provide a simple example motivating the need of multiple hypothesis test correction. Let us consider 100 experts that are requested to provide a binary decision on a given problem (for example a prediction about the increase (state 1) or decrease (state 0) of the price of a financial asset in a given time horizon). This activity is repeated for 10 distinct time intervals and an observer aims to evaluate the performance of experts with respect to a null hypothesis assuming that experts are just providing random predictions. Under the null hypothesis the probability $p_8$ that a specific expert is providing at least 8 correct predictions is
\begin{equation}
p_8=\frac{\binom{10}{10}+\binom{10}{9}+\binom{10}{9}}{2^{10}}=0.0547
\end{equation} 
However, such a low probability does not prevent us to observe a rejection of the null hypothesis in the panel of 100 experts just due to the fact that we are simultaneously performing a test on 100 distinct experts. In fact, under the random null hypothesis, the probability $p$ to observe at least one expert in the set of 100 making at least 8 correct prediction is 
\begin{equation}
p=1-(1-p_8)^{100}=0.9964
\end{equation} 
Therefore to avoid false positive due to the large number of performed tests a multiple hypothesis test correction is needed.
The most restrictive multiple hypothesis test correction is the so-called Bonferroni correction  \cite{Miller1981}. When a statistical threshold $\alpha$ is chosen to perform a single statistical test, the Bonferroni correction is done by using $\alpha_B=\alpha/N_t$ as a statistical threshold for the single test, where $N_t$ is the number of test performed over the entire network. The Bonferroni correction increases the statistical precision (by minimizing the number of false positive of the test) but decreases the accuracy of the estimation because the severe reduction of the statistical threshold can be associated with the presence of a large number of false negative. 

In addition to the Bonferroni correction several other multiple hypothesis test correction approaches have been proposed. Among them the one that is currently most used is the so-called control of the false discovery rate (FDR) \cite{Benjamini1995}.  The control of the FDR works as follows: the $p$-values of distinct tests are first arranged in increasing order ($p_1<p_2<...<p_t$) and then a threshold for the rejection of the null hypothesis is obtained 
by considering the largest $t_{max}$ such that $p_{t_{max}}<t_{max} ~ \theta_B$. 

\section{Statistically validated networks}
Bipartite networks are quite common in complex systems. In a bipartite network nodes of one set (say set A) are connected with nodes of the second set (say set B) but no connection is observed among any pair of nodes of set A and among any pair of nodes of set B. A classic example of bipartite network is the actor-movie network where each actor is linked with movies where he or she played, whilst there is no connection amongst movies nor amongst actors.  In most cases bipartite networks have been investigated by performing the so-called projection in one of the two sets (for example the projection of the network of actors having played at least once in the same movie.

For a projected network obtained from a bipartite network a basic null hypothesis for each link of the projected network can be formulated by assuming random connections between nodes of set A and nodes of set B that are preserving the degree of A and B nodes. The detection of a statistically validated network works as follows \cite{Tumminello2011}. Let us consider a bipartite system where links connect the $N_A$ elements of set A (e.g. actors) to the $N_B$ elements of set B (e.g. movies). In the present description, we focus on the projected network of nodes of set A but the same approach can also be applied to the projected network of nodes of set B. The projected network of nodes of set A is obtained by linking together those nodes of A with at least one common first neighbor element of B. The method  statistically validates each link of the projected network against a null hypothesis of random co-occurrence of common neighbors that takes into account the degree heterogeneity of elements of set A.  For each pair of elements $i$ and $j$ of set A, we consider $N_{i,j}$, i.e. the number of common neighbors in set B. Let us label as $N_i$ and $N_j$ the degree of node $i$ and $j$ respectively. Under the hypothesis that elements $i$ and $j$ randomly connect to the elements of set B, the probability that elements $i$ and $j$ share $X$ neighbors in set B is well approximated \footnote{The probability is exact when the set B does not present degree heterogeneity, i.e. when the degree of all nodes is the same. However, numerical simulations have shown that in most cases the approximation is quite good  also in the presence of strong degree heterogeneity in the set B. For an approach controlling the heterogeneity of set B see \cite{Tumminello2011}} by the hypergeometric distribution
\begin{equation}
\label{hypertheor}
H(X|N_B,N_i,N_j)=\frac{{N_i \choose X} {N_B-N_i \choose N_j-X}}{{N_B\choose N_j}}.
\end{equation}
The probability obtained in Eq.~(\ref{hypertheor}) allows to estimate a \emph{p-value} $p(N_{i,j})$ to the empirical observation of $N_{i,j}$ neighbors  or more between node $i$ and $j$. The  \emph{p-value} is
\begin{equation}
\label{pvaltheor}
p(N_{i,j})=1-\sum_{X=0}^{N_{i,j}-1}H(X|N_B,N_i,N_j).
\end{equation}
With this approach, by performing a statistical test for each pair of nodes $i$ and $j$ one can associate a \emph{p-value} to all links of the projected network of nodes of set A.

The number of tests to be performed is given by the number of links present in the projected network. This number is usually very high and for this reason a multiple hypothesis test correction is necessary to avoid a large number of false positive. Therefore after computing \emph{p-values} associated with all pairs of nodes of set A the rejection of the null hypothesis is verified for each link under a chosen multiple hypothesis test correction (for example a Bonferroni or a FDR correction). The links that reject the null hypothesis are therefore included in the statistically validated network.

Undirected statistically validated networks have been detected in actor-movie network \cite{Tumminello2011}, in gene-genome network \cite{Tumminello2011}, in investor-investment/decision network \cite{Tumminello2012}, in caller-receiver mobile phone subscriber \cite{Li2014a,Li2014b}, in bank/lending-bank/borrowing network \cite{Hatzopoulos}, in lead-lag return of financial assets \cite{Curme} and in lead-lag inventory prediction in the foreign exchange market \cite{Challet}.

One prominent aspect of a statistically validated network is its ability to detect under-expressed occurrences between a pair of nodes. In fact the test described above for over-expression of $N_{ij}$ (with respect to the null hypothesis) can also be performed to detect under-expression of the same quantity. It should be noted that this is a quite unique opportunity to detect  avoided relationships in a projected network. In fact, avoided relationships are unknown in projected network obtained from a bipartite network. The detection of avoided relationships was first performed in \cite{Hatzopoulos}. It is performed by considering a two tail statistical test and the \emph{p-value} associated with the under expression of $N_{i,j}$ is given by
\begin{equation}
\label{pvaltheor}
p(N_{i,j})=\sum_{X=0}^{N_{i,j}}H(X|N_B,N_i,N_j).
\end{equation}
It is worth noting that in this case also the absence of co-occurrence between nodes $i$ and $j$ can be informative. For this reason the number of tests to be performed can be much higher than in the case of the one tail test. Specifically, for the two tails test the number of tests is of the order of $N_A^2$, where $N_A$ is the number of nodes of set A. This implies that usually the power of the test for both under-expression and over-expression is lower than in the case of a test performed only for over-expressions.
\begin{center}
\begin{figure}
\includegraphics[scale=0.7]{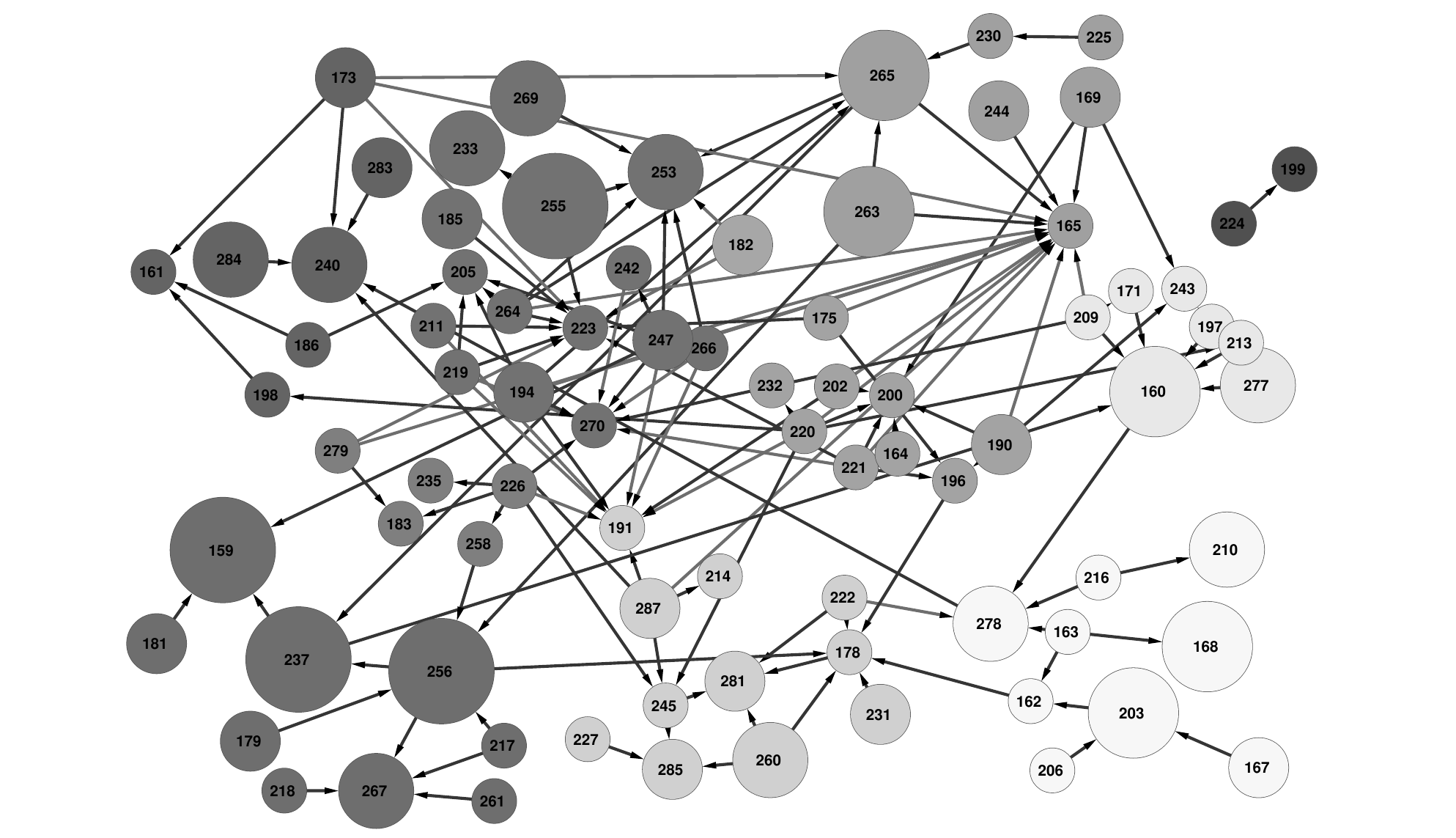}
\caption{Statistically validated network (with the Bonferroni multiple hypothesis test correction) of the Italian segment of the e-MID market during the maintenance period from 10 September 2008 to 9 December 2008. Arrows originate from the lender aggressor. The different gray levels indicate the node membership obtained by applying a community detection algorithm. Light gray links are under-expressed links, while black links are over-expressed ones. Adapted from \cite{Hatzopoulos}.}
\label{figA}
\end{figure}
\end{center}
In Fig. \ref{figA}  we show the statistically validated network of the Italian segment of the e-MID market  obtained in  \cite{Hatzopoulos} from data of the maintenance period from 10 September 2008 to 9 December 2008. Each node indicates a bank. Identity of the bank was not provided by the seller of the database for confidentiality reasons. Links (both originally and after statistical validation) are directed and the direction indicates the lender aggressor (i.e. the bank deciding to close the credit contract while acting as a lender). Links are drawn with two levels of gray. Specifically, over-expressed links are black links whereas under-expressed ones are light gray links. Both types of links are detected in this system manifesting that networked relationships occurs in this market. It is worth noting that the method highlight something that it is not typically encoded in traditional projected network. This is the information concerning under-expression of relationships. For example the observation that bank labeled as 165 is involved in loans less than expected in terms of its trading activity with many other banks when it acts as a borrower could be highly informative. Equally informative could be the fact that  a few banks (labeled as 169, 244, 263 and 265) take an opposite role by showing over-expressed links with the same bank.

\section{Examples of applications of statistically validated networks}
Statistically validated networks highlight whether the number of repeated events observed between pairs of nodes of a set of a bipartite network is compatible or not with a null hypothesis that it is assuming heterogeneity of action of the different nodes. Heterogeneity is a ubiquitous aspect of complex systems. Its detection is very important but it is just the first characterization of the system. The next step after the detection of heterogeneity (done, for example, by detecting a leptokurtic degree distribution) is the detection of  links whose co-occurrences are not compatible with the null hypothesis.

Mobile phone communications have been investigated with tools of network science since many years \cite{Onnela}. Pre-processing of the data is extremely important in this type of investigation because many business and technical activities are performed in the communication system in addition to the social activities of exchange and diffusion of information. In reference \cite{Li2014b} it was shown that the study of statistically validated networks makes much more robust the detection of communication motifs that can be safely attributed to social contacts. In other words, statistically validated network are more resilient than ordinary networks to errors, or sets of links originating from activities not primarily related with the main activity of the considered complex systems as, for example in the considered case, phone calls due to contacts with call centers and/or phone calls scheduled to promote products or services. This type of collateral activities often present in a complex system are typically difficult to be disentangled from the main type of activity performed on it. In these cases, statistically validated networks of the system naturally focus on the main aspects of the considered complex system.

Another area of application concerns extremely dense complex networks. It is known that the investigation of the mesoscopic structures in complex networks is more problematic in highly dense networks. In fact, community detection algorithms are often not able to produce node's partitions when the density of link is very high. In these cases, detecting statistically validated networks help in the detection of mesoscopic structures of the network. Such type of application of the statistically validated networks has been performed in the study of the bipartite network crimes-criminal suspects \cite{Tumminello2013} and in the detection of clusters of investors characterized by the same investment profile \cite{Tumminello2012,Musciotto2016,Musciotto2018}.

Statistically validated networks can also be used to detect preferential or avoided networked relationships in a financial market. Several financial markets are anonymous and/or have policy incentives to allow equal access to transactions to all participants also in the presence of several sources of heterogeneity of financial actors. It is therefore of importance to detect networked relations that cannot be explained by a random null hypothesis of transaction matching taking into account the heterogeneity of the different financial actors. Statistically validated networks have been observed in a study of the Italian Interbank market \cite{Hatzopoulos} showing that networked preferential and avoided interactions are present in the Interbank market.

\section{Community detection in statistically validated networks}
In the previous section we have seen that the study of statistically validated networks can be useful to detect clusters of nodes (i.e. communities) in highly dense networks where ordinary community detection algorithms may fail. In ref. \cite{BongiornoCore}, authors show that the investigation of statistically validated network is able to detect cores (i.e. the most internal subset of given partitions of nodes)  of communities. By using a bipartite benchmark synthetically generated authors were able to show that the accuracy and precision of the partition of nodes of projected networks obtained with respect to the reference partition is high when it is measured in terms of the adjusted Rand index and the adjusted Wallace index. In fact, the detection of cores of nodes is always highly precise, although the accuracy of the methodology can be limited in some cases.
 
\begin{center}
\begin{figure}
\includegraphics[scale=0.45]{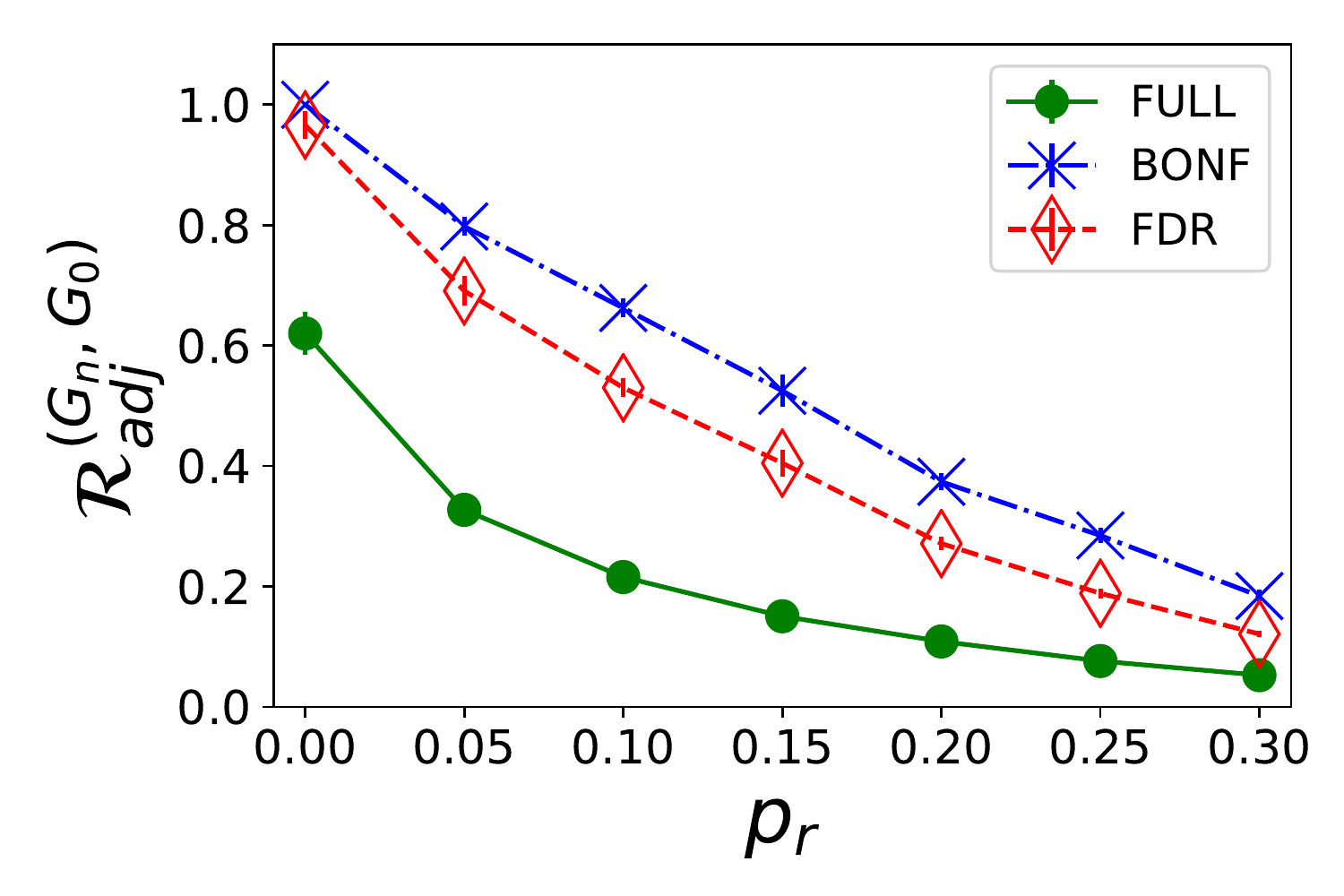}
\includegraphics[scale=0.45]{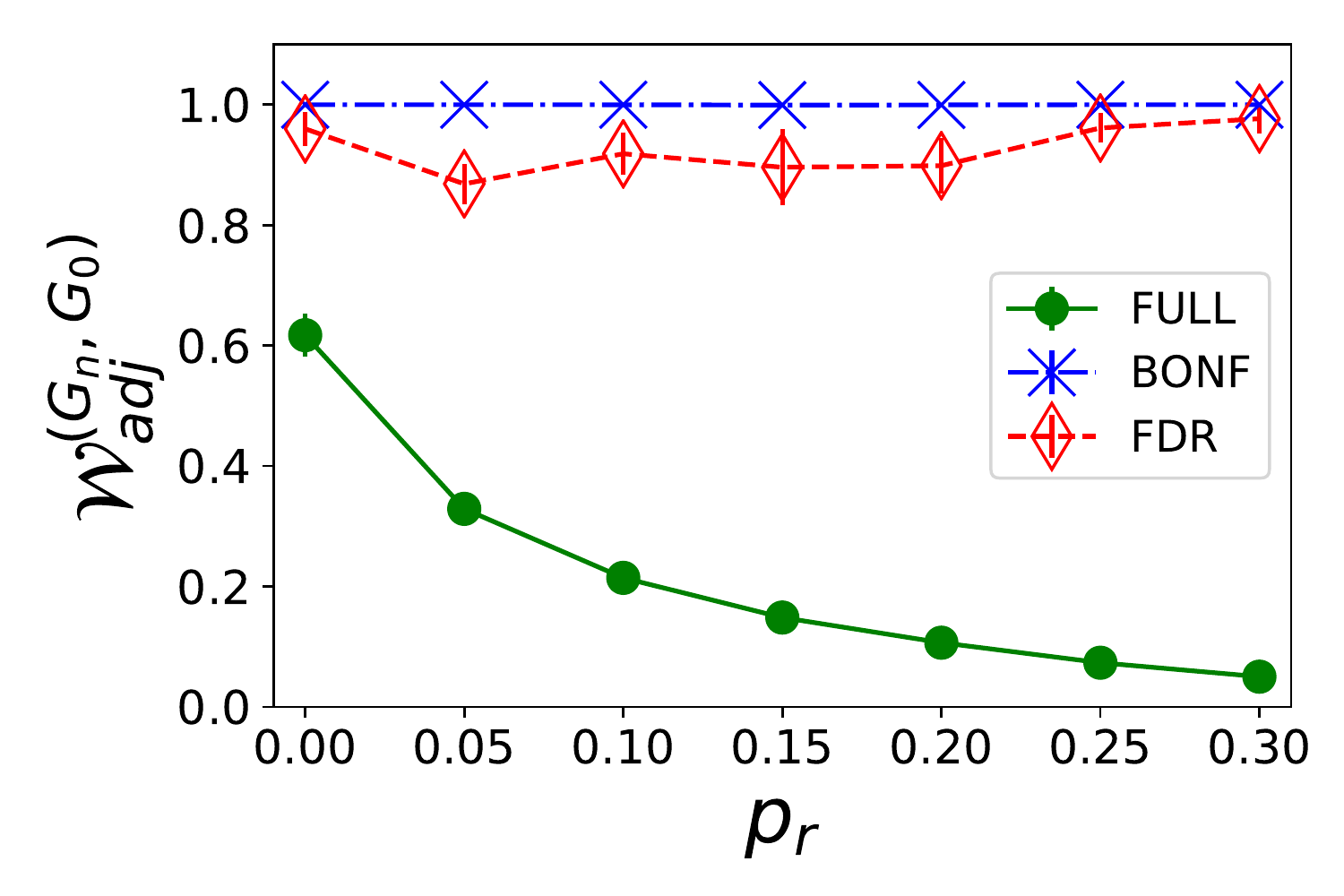}
\caption{Coauthorship database. Left panel: Average $R_{adj}$ value between 100 partitions of the original projected network (green circles), the Bonferroni statistically validated network (blue crosses), and FDR statistically validated networks (red diamonds) obtained as different stochastic realizations for each investigated value of $p_r$ and the best partition $G_0$ obtained in the absence of noise. The standard deviation of each value is of the order of the symboi. Right panel: Average $W_{adj}$ of the same partitions. Reproduced from \cite{BongiornoCore}.}
\label{fig1}
\end{figure}
\end{center}

It is worth noting that the authors were able to show that the detected cores of communities are highly  robust with respect to the presence of errors or missing entries in the investigated bipartite network. This conclusion was reached by adding noise to the benchmark and/or by pruning to a certain degree the bipartite network. The analysis was also performed in widely investigated real database as the co-autorship network originally investigated in \cite{Newman2001}. In Fig. \ref{fig1} we show the results they found when comparing partitions obtained with (i) the original projected network of authors (called by the authors the "full" network),(ii) the statistically validate network obtained with the multiple hypothesis test correction of the control of the FDR and (iii) the statistically validate network obtained with the Bonferroni multiple hypothesis test correction.
In the figure, different values of $p_r$ refer to a different level of rewiring of links of the bipartite network. As usual, the rewiring was performed between pairs of nodes with the same degree. The different partitions obtained for different values of $p_r$ were compared with the best partition $G_0$ obtained from the original projected network by using the Louvain algorithm \cite{Blondel}. The Louvain algorithm was also used in the detection of communities of all networks. 

In Fig. \ref{fig1} each point observed for each value of $p_r$ is the average value obtained by computing 100 different realizations. The metric used for the comparison with the $G_0$ partition are the adjusted RAND index $\it{R}_{adj}$ and the adjusted Wallace index  $\it{W}_{adj}$. The adjusted RAND index \cite{Hubert} is obtained by considering true positive, false positive, true negative and false negative of pairs of nodes in a given partition compared with a reference partition (i.e. the statistical accuracy of pair classification) whereas the adjusted Wallace index \cite{BongiornoCore} measures how many pairs of nodes that belong to the same community are indeed in the same community also  in the reference partition (i.e. the statistical precision of pair classification). Left panel of Fig. \ref{fig1} show the adjusted RAND index the best partition of the original network and 100 partitions obtained for the same network for each value of $p_r$ ranging from 0.05 to 0.3. The panel  also shows $R_{adj}$ for statistically validated network obtained with the control of FDR correction.  The partitions obtained by using FDR statistically validated networks are always more robust to noise than those obtained by investigation the original  network. In the right panel of Fig. \ref{fig1}, authors show $W_{adj}$ for the same networks. The fact that the adjusted Wallace index is always very close to one show that cores of communities detected by investigating the FDR statistically validated networks decrease in similarity (i.e., $R_{adj}$ values) with the partition $G_0$ not due to a decrease in precision but rather a decrease in accuracy (i.e. several pairs of nodes are not included in the statistically validated networks but those included still are clustered accordingly to $G_0$). In fact, $W_adj$ of FDR statistically validated networks does not go below 0.85 for all values of $p_r$ , whereas authors observe values of $W_{adj}$ as low as 0.1 for partitions obtained from the original network when $p_r = 0.3$. The difference are even more marked when the performance of the original network is compared with the performance of the Bonferroni network (blue symbols). The Bonferroni network provides more precise results at the cost of decreasing their statistical accuracy. In summary, the informativeness of the detected cores of communities is robust with respect to noise  added or present in the database.
The investigation of communities detected in statistically validated networks can therefore be highly informative when researchers are interested to highlight those nodes constituting the most stable subsets of the clusters of nodes present in a complex system.  

\section{Software for the computation and analysis of statistically validated networks}
A few software packages have been written to perform statistical validation of links in complex network. A code written in R is downloadable at the webpage:\\
http://mathfinance.sns.it/statistical\_validation/\\
A package written in R will soon be submitted to the CRAN repository \cite{ChalletPersonal}. Another software resource can be found at GhitHub at the webpage:\\
https://github.com/cbongiorno/Bipartite-Tools\\
This last program is a software tool for testing community detection algorithms on bipartite projected networks and also contains a module to detect a statistically validated network in bipartite networks. The software module {\it validate} contains the functions for statistically validated networks. The software module {\it metrics} contains other useful functions (for examle a function to estimate the adjusted Wallace index between a partition and a reference partition. The code includes a wrapper to the Louvain community detection method written by E. Lefebvre and released under GNU Licence.

\section{Conclusions} 
Network description of complex systems is a very rich description. For this reason a network can usually be analyzed at different hierarchical levels. In this contribution we have considered the analysis of network at the level of a single link. The selection of specific nodes and links is done by performing statistical tests of nodes and links against a chosen null hypothesis. We believe that this method is highly informative when the null hypothesis is taking into account the heterogeneity empirically observed in the investigated complex system. Specifically, in the present contribution we have discussed the so-called disparity filter and the method for the detection of statistically validated networks in bipartite systems. Statistically validated networks can be used for (i) highlighting over-expressed or under-expressed pair relationships between nodes of the investigated system. (ii) pre-processing of large datasets where relationships between subsets of pair of nodes can be attributed to idiosyncratic motivations that are different from the most common motivations observed in the investigated system, and (iii) detection of the cores of communities of nodes present in the projected network of a bipartite network. In summary, a process of statistical validation systematically performed on all links of the network can highlight a set of links that are due to relationships that cannot be due only to the heterogeneity of the nodes. These over-expressions are often highly informative about the nature of the system as, for example, in the case of networked structures detected in the transaction network observed in a financial market. The detection of under-expressions is also conveying important information about the system and the importance of this information is hard to overestimate because this type of information is hardly observable with classic approaches of network science.

\acknowledgments
We would like to thank the many colleagues with whom we have introduced statistically validated networks and applied them to several complex systems. In particular we are grateful to our co-authors: H. Aoyama, C. Bongiorno, C. Curme, C. Edling, Y. Fujiwara,  M. Gallegati, G. Gurtner, V. Hatzopoulos, H. Iyetomi, Z.Q. Jiang, K. Kaski, D.Y. Kenett, J. K\'ertesz, M.X. Li, F. Lillo, F. Liljeros, L. Marotta, F. Musciotto, V. Palchykov, J. Piilo, J. Sarnecki, H.E. Stanley, M. Tumminello,  W. J. Xie, and W.X. Zhou for fruitful discussions shared during the joint work.

\end{document}